\documentstyle[aps,12pt,epsfig,amssymb]{revtex}

\newcommand{\bm}{\bibitem}

\newcommand{\cp}{\chi^{(+)}}
\newcommand{\cm}{\chi^{(-)*}}

\newcommand{\vv}{V_{bc}({\bf r}_1)}
\newcommand{\ri}{{\bf r}_i}
\newcommand{\ro}{{\bf r}_1}
\newcommand{\ak}{{\bf k}_a}
\newcommand{\bq}{{\bf k}_b}

\newcommand{\rc}{{\bf r}_c}
\newcommand{\cq}{{\bf k}_c}
\newcommand{\we}{\Psi^{(+)}_a(\xi_a,{\bf r}_1,{\bf r}_i)}

\newcommand{\fa}{ _2F_1(1-i\eta_a,1-i\eta_b;2;D(0))}
\newcommand{\fB}{ _2F_1(-i\eta_a,-i\eta_b;1;D(0))}

\tighten
\begin{document}

\draft

\title{Core excitation in Coulomb breakup reactions}
\date{\today}
\author{R. Shyam$^{a,b}$ and P. Danielewicz$^a$}
\address {
$^a$
National Superconducting Cyclotron Laboratory and\\
Department of Physics and Astronomy, Michigan State University,
\\
East Lansing, Michigan 48824, USA\\
$^b$ Saha Institute of Nuclear Physics, Calcutta 700064, India}
\maketitle

\begin{abstract}
Within the pure Coulomb breakup mechanism,
we investigate the one-neutron removal reaction of the type
A(a,b$\gamma$)X with $^{11}$Be and $^{19}$C projectiles
on a heavy target nucleus $^{208}$Pb at the beam energy of 60
MeV/nucleon.  Our intention is to examine
the prospective of using these reactions to study the structure of
neutron rich nuclei.  Integrated partial cross sections
and momentum distributions for the
ground as well as excited bound states of core nuclei
are calculated within the finite range distorted wave Born
approximation as well as within the adiabatic model of the
Coulomb breakup.
Our results are compared with those obtained in the studies of
the reactions on a light target where the breakup proceeds via
the pure nuclear mechanism.  We find that
the transitions to excited states of the core are quite weak
in the Coulomb dominated
process
as compared
to the pure nuclear breakup.
\end{abstract}
\newpage
\section{Introduction}

The single-nucleon transfer reactions, induced by light
as well as heavy ions, have been established as a useful
tool in probing the single-particle components of the wave functions
of stable nuclei (see e.g.\ \cite{sat83,aus70,fes92,gle83}).
The theory of these reactions developed within the framework of
the distorted wave Born approximation (DWBA) \cite{sat64} has been
widely used to analyze the absolute magnitudes and shapes of
measured cross sections and to deduce the structure information
including
angular momentum assignments, occupation probabilities and spectroscopic
factors of the ground as well as excited states of the residual nuclei.

Nonetheless, transfer reactions are not yet routinely used
in
probing the structure of exotic nuclei near the neutron and proton
drip lines, even though the first theoretical feasibility study
\cite{len98}
for such investigations with transfer reactions and the first
experimental results \cite{win00} for the $^{11}$Be(p,d)$^{10}$Be reaction
have been already reported. With the currently available
experimental techniques, the measurements of these reactions
involving drip line nuclei are performed
in the inverse kinematics with low intensity projectile beams.
This puts severe experimental restrictions as the corresponding
cross sections are usually low. Furthermore, the theoretical analysis
of these data in terms of the DWBA gets complicated as the usual
well-depth search method to calculate the wave function of the
transfered particle becomes unreliable \cite{win00}, and the methods
such as Skyrme Hartree-Fock theory need to be invoked \cite{len98} for a
proper description of these wave functions.

Recently, an alternative new and more versatile technique for
investigating the spectroscopy of nuclei near the drip line has been
developed \cite{nav98,aum00,nav00,val00}.  In this method, referred to
as the (a,b$\gamma$) reaction in the following, one nucleon
(usually the valence or halo) is removed from the projectile (a) in its
breakup reaction within the field of a target nucleus.
The states of the core fragment (b) populated in this reaction are
identified by their gamma ($\gamma$) decay. The $\gamma$-ray intensities
are used to determine the partial breakup cross sections to
different core states. The signatures of the orbital angular momentum
$\ell$ associated with the relative motion of core states with
respect to the valence nucleon (removed from the projectile)
are provided by the measured parallel momentum distributions
\cite{han96}.

This method improves the experimental conditions for working with
projectiles of low beam intensities because of:
(i) large partial cross sections for transitions to various bound
states of the core fragment, even in experiments done with
high-energy projectiles, (ii) possibility of using thick targets,
and (iii) strong forward focusing. These features may be contrasted with
those of the corresponding transfer reactions. In addition, while, in the
case of transfer reactions, the angular distributions of the
ejectile lose their characteristic $\ell$-dependence at high energies
\cite{bou81}, the longitudinal momentum distributions of the core states
in the breakup reactions continue to show a strong dependence on $\ell$.

Most of the studies of the (a,b$\gamma$) reaction
performed so far involve a light $^9$Be target, where the breakup
process is governed almost entirely by only the nuclear interaction
between the projectile fragments and the target. Since, this reaction
is essentially inclusive in nature (as the measurements are performed
only for the heavy core fragment), the nuclear partial cross sections have
contributions from both elastic (also known as diffraction dissociation)
and inelastic (also known as stripping or breakup-fusion) breakup modes
\cite{baur00,kasa82}. Several attempts have been made to calculate the
elastic and inelastic nuclear breakup cross sections of halo nuclei and
they were either based on the semiclassical methods \cite{angela} or on
the eikonal approximation \cite{yab92,ber96,bar96,par00}. The
fragment-target interactions are dealt with differently in these two
approaches which could be important for the light targets \cite{anb00}.
Data of Refs.\ \cite{nav98,aum00,nav00,val00} have been analyzed in terms
of an eikonal model \cite{tos99} with core-target and neutron-target
interactions treated in the black disc approximation and in the optical
limit of the Glauber theory, respectively. In order to extract
unambiguous spectroscopic informations from the (a,b$\gamma$)
type of measurements performed on a light target, it is quite desirable
to develop the calculations of nuclear breakup reactions within the DWBA
theory as has been done
for the breakup of stable projectiles \cite{bau84,mac84}.

However, currently a full quantum mechanical theory of the pure
Coulomb breakup reaction, formulated within the framework of
the post-form distorted-wave Born-approximation, is well
established and has been applied successfully to investigate the breakup
of halo nuclei \cite{cha00}. Finite range effects are accounted for
in this theory which can be applied to projectiles of any ground-state
orbital angular-momentum structure.
Moreover, an alternative theory of the Coulomb breakup
reactions within the framework of an adiabatic model has also
been formulated \cite{tos98}. The expressions for the breakup amplitude
within this theory are very similar to those of the
finite-range DWBA
theory, although the two have been derived under quite different
assumptions. In the adiabatic model, it is assumed that the excited states
of the projectile are degenerate with the ground state.
In the studies of the breakup reactions done so far (where the core fragments
were assumed to remain in their ground states), the two theories
produced almost identical results \cite{cha00}. However, with the
excitation of the core, the one-neutron separation energies
increase significantly. It would,
therefore, be interesting to see if the two models lead to
different results in these cases.

There are no adjustable parameter in either of the theories of pure
Coulomb breakup reaction.  Assuming that the processes, in which the
mutual excitation of the target nucleus takes place due to the Coulomb
interaction, contribute negligibly, the inelastic breakup mode is absent
in the pure Coulomb breakup reactions.  This is an added advantage as
there is some ambiguity regarding the calculation of this mode which
dominates the partial cross sections for the excited core states in the
nuclear breakup process \cite {tos99}.  

In this paper, we present calculations of the pure Coulomb breakup
contributions to the partial cross sections and longitudinal momentum
distributions of the ground as well as excited states of the
core fragments, $^{10}$Be and $^{18}$C, in the (a,b$\gamma$) type
of reaction induced by $^{11}$Be and $^{19}$C projectiles, respectively
on a $^{208}$Pb target at the beam energy of 60 MeV/nucleon. We assume
that the states
of the core fragments are the same as those seen in the similar reactions
studied on the $^9$Be target.  Our aim is to determine if there
are quantitative differences in the ${\it relative}$ populations
of the core states in the pure Coulomb breakup mechanism,
as compared to those observed in the pure nuclear breakup process.
We shall also look whether there are
differences in the predictions of the finite range DWBA and adiabatic models
of the breakup reactions leading to the core excited states.

We want to make it clear from the very beginning that it is not our 
intention  
to imply that the nuclear breakup contributions are negligible
for the reactions investigated by us. Our results should be viewed as
complementing contributions from
the nuclear breakup process; in any complete theory
both contributions must be considered on an equal footing.

In the next section we briefly present the formalism of the Coulomb
breakup reactions. The results of our calculations and discussions
are presented in section III. A summary and the conclusions of our
work are given in section IV.

\section{Formalism}

We consider the reaction $ a + t \rightarrow b + c + t $, where the
projectile $a$ breaks up into fragments $b$ (charged)
and $c$ (uncharged) in the Coulomb
field of a target $t$. The chosen coordinate system is shown in Fig. 1.
\begin{figure}[here]
\begin{center}
\mbox{\epsfig{file=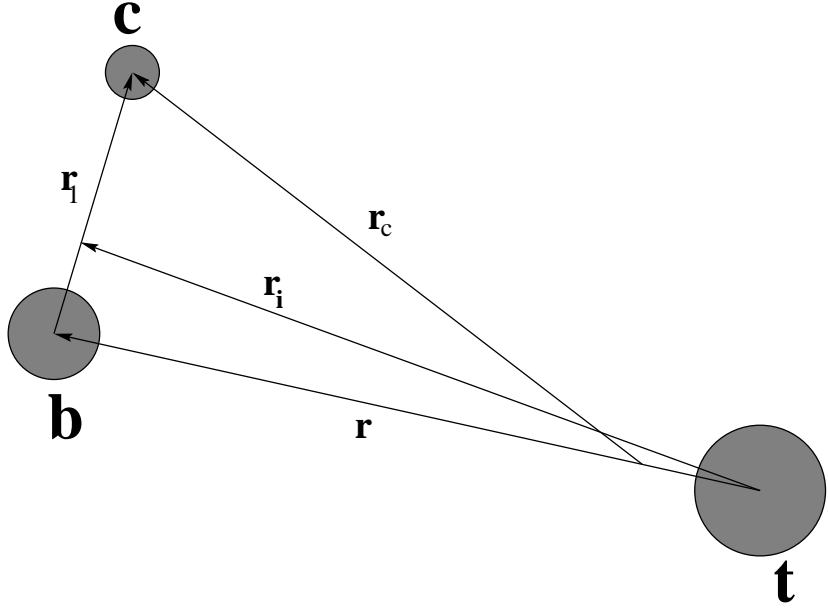,height=8.0cm}}
\end{center}
\caption{
The three-body coordinate system. The charged core, valence
neutron and target are denoted by $b$, $c$ and $t$, respectively.
}
\label{fig:figa)}
\end{figure}
\noindent
The position vectors satisfy the following relations:
\begin{eqnarray}
{\bf r} &=& \ri - \alpha\ro,~~ \alpha = {m_c\over {m_c+m_b}} \,
,   \\
\rc &=& \gamma\ro +\delta\ri, ~~ \delta = {m_t\over {m_b+m_t}},
~~  \gamma = (1 - \alpha\delta) \, .
\end{eqnarray}

The starting point of both the finite-range distorted-wave Born
approximation
(FRDWBA) and of the adiabatic model of the Coulomb breakup is
the post-form $T$-matrix of the reaction given by
\begin{eqnarray}
&&T = \int  d\xi d\ro d\ri \cm_b(\bq,{\bf r})\Phi^*_b(\xi_b)
\cm_c(\cq,{\bf \rc})\Phi^*_c(\xi_c)\vv \we.
\end{eqnarray}
The functions $\chi$ are the distorted waves for the relative
motions of
$b$ and $c$ with respect to $t$ and the center of mass (c.m.)\
of the $b+t$ system, respectively.  The functions $\Phi$ are
the internal state wave functions of the concerned particles
dependent on the internal coordinates $\xi$. The function $\we$
is
the exact three-body scattering wave function of the projectile
with a wave vector $\ak$ satisfying outgoing boundary
conditions.  The vectors $\bq$ and $\cq$ are the Jacobi wave
vectors of $b$
and $c$, respectively, in the final channel of the reaction.
The function $\vv$ represents the
interaction between $b$ and $c$.  As we concentrate only on the
pure Coulomb
breakup, the function $\chi^{(-)}_b({\bq},{\bf r})$ is taken as
the Coulomb
distorted wave (for a point Coulomb interaction between
the charged core b and the target) satisfying incoming wave
boundary conditions, and
the function $\chi^{(-)}_c({\cq},{\rc})$ is just a plane wave
as there is no Coulomb interaction between the target and the
neutral fragment $c$.

In the distorted wave Born approximation (DWBA), we write
\begin{eqnarray}
\we = \Phi_a(\xi_a,\ro)\cp_a(\ak,\ri),
\end{eqnarray}
The assumption inherent in Eq.~(4) is that the breakup channels
are very weakly coupled and hence this coupling needs to be treated only
in the first order. In this equation 
the dependence of $\Phi_a$ on ${\bf r}_1$
describes the relative motion of the fragments $b$ and
$c$ in the ground state of the projectile.
The function $\cp_a(\ak,\ri)$ is the Coulomb distorted
scattering wave describing the relative motion of the c.m.\ of the
projectile with respect to the target, satisfying outgoing wave boundary
conditions. It may be noted that the particular case of the pure 
Coulomb breakup of a projectile involving one uncharged fragment, 
where the choice of the coordinate {\bf r} may appear more natural
to describe the relative motion between the projectile and the target,
follows from this expression as discussed below. 

The integration over the internal coordinates $\xi$ in the
$T$-matrix gives
\begin{eqnarray}
\int d\xi\Phi^*_b(\xi_b)\Phi^*_c(\xi_c)\Phi_a(\xi_a,\ro)
& = & \sum_{\ell mj\mu} \langle \ell mj_c\mu_c|j\mu\rangle
 \langle j_b\mu_bj\mu|j_a\mu_a\rangle i^\ell \Phi_a(\ro),
\end{eqnarray}
with
\begin{eqnarray}
\Phi_a(\ro) & = & u_{\ell}(r_1) Y_{\ell m}({\hat{\bf r}}_1).
\end{eqnarray}
In Eq.~(6),  $\ell$  (the
orbital angular momentum for the relative motion between fragments
$b$ and $c$) is coupled to the spin of $c$ and
the resultant channel spin $j$ is coupled to the spin
$j_b$ of the core $b$ to yield the spin of $a$ ($j_a$).
The $T$-matrix can now be be written as
\begin{eqnarray}
T & = & \sum_{\ell mj\mu} \langle \ell mj_c\mu_c|j\mu\rangle
\langle j_b\mu_bj\mu|j_a\mu_a\rangle i^\ell
\hat{\ell}\beta_{\ell m}(\bq,\cq;\ak),
\end{eqnarray}
where
\begin{eqnarray}
&&\hat{\ell}\beta_{\ell m}(\bq,\cq;\ak)  =
\int d\ro d\ri\cm_b(\bq,{\bf r})e^{-i\cq.\rc} \vv u_\ell (r_1) Y_{\ell m}
({\hat r}_1)\cp_a(\ak,\ri).
\end{eqnarray}
with $\beta_{\ell m}$ being the reduced $T$-matrix and
with ${\hat \ell} \equiv \sqrt{2\ell + 1}$.

Equation (8) involves a six dimensional integral which makes the
computation of $\beta_{\ell m}$ quite complicated.
The problem gets further acute because
the integrand has a product of three scattering waves
that exhibit an oscillatory behavior asymptotically.
Therefore, approximate methods have been used, such as the zero
range approximation (ZRA) (see
e.g.\ \cite{sat83,aus70,gle83}), in which the product
$\vv\Phi_a(\ro)$ is replaced by a delta function, or
the Baur-Trautmann approximation \cite{btr72},
where the projectile c.m.\ coordinate is replaced by that of
the core-target system (i.e. $\ri \approx {\bf r}$).
Both these approximations lead to a
factorization
of the reduced amplitude into two independent parts, which
reduces the computational complexity. However, the application
of both these methods to the  reactions of halo nuclei is
questionable \cite{cha00}. The ZRA necessarily
restricts the relative motion between $b$ and $c$ in the projectile to
$s$-state only. Even for such cases, this approximation may not be 
valid for heavier projectiles and at higher beam energies (see 
e.g. \cite{shyam85}). The Baur-Trautmann approximation is justified
if the c.m. of the $b$+$c$ system is shifted towards $b$ (which is 
indeed the case if $m_b \gg m_c$). However, since {\bf r}$_i$
occurs in association with the wave vector 
{\bf k}$_a$, whose magnitude is quite appreciable at the higher 
beam energies, the neglected piece of 
{\bf r}$_i$ (i.e. $\alpha$ {\bf r}$_1$) may still contribute substantially.

In the FRDWBA theory, the Coulomb distorted wave of particle $b$ in
the final channel is written as \cite{cha00}
\begin{eqnarray}
\chi^{(-)}_b(\bq,{\bf r}) & = & e^{-i\alpha{\bf K}.\ro}
                           \chi^{(-)}_b(\bq,\ri).
\end{eqnarray}
Equation (9) represents an exact Taylor series expansion about
${\bf r}_i$ if
$ {\bf K} = -i\nabla_{{\bf r}_i}$ is treated exactly. However,
instead of doing this we employ a local momentum approximation
\cite{shyam85,braun74a}, where the magnitude of momentum ${\bf K}$
is taken to be
\begin{eqnarray}
K(R) & = &{\sqrt {{2m\over \hbar^2}(E - V(R))}}.
\end{eqnarray}
Here $m$ is the reduced mass of the $b-t$ system,
$E$ is the energy of particle $b$ relative to the target in the
c.m.\ system and $V(R)$ is the  Coulomb potential
between $b$ and the target separated by $R$. Thus,
the magnitude of the momentum of ${{\bf K}}$ is evaluated at
some separation
$R$ which is held fixed for all the values of ${r}$. The value
of $R$ was taken to be equal to 10 fm. For reactions under investigation
in this paper,
the magnitude of $K$ remains constant for distances larger than
10 fm \cite{cha00}. Due to the peripheral nature of the breakup reaction,
the region $R \gtrsim 10$ fm contributes maximum to the cross
section.
In fact, the calculated cross sections change by only about 5$\%$ if
$R$ is varied between 5 to 10 fm and with a further increase in R
the change is less than 1$\%$. Furthermore, the results of the
calculations for these reactions, at the beam energies under
investigation, are almost independent
of the choice of the direction of momentum ${\bf K}$\cite{cha00}.
Therefore, we have taken the directions of ${{\bf K}}$ and
${\bf k_b}$ to be the same in all the calculations presented in this
paper. It may be remarked here that in
Ref.~\cite{ban98} an approximation similar to Eq.~(9) was
applied
to the Coulomb distorted wave of the incident channel.  That
procedure
brings in two difficulties. Firstly, the choice of the
direction of the local momentum is somewhat complicated as
directions of
the both fragments in the final channel will have to be brought
into
consideration. Secondly, the procedure may produce a deviation
from the exact DWBA approximation.

On substituting Eq.~(9) into Eq.~(8), we obtain the following
factorized form of the reduced amplitude
\begin{eqnarray}
{\hat \ell}\beta^{FRDWBA} _{\ell m} & = & \left [
\int d\ro e^{-i(\gamma\cq - \alpha {\bf K}).\ro}V_{bc}(\ro) u_\ell(\ro)
Y_{\ell m}({\hat r}_1)\right ] \nonumber \\
& \times & \left [\int d\ri \chi^{(-)*}_b(\bq,\ri) e^{-i\delta\cq.\ri}
\cp_a(\ak,\ri) \right ].
\end{eqnarray}
This amplitude differs from those in earlier studies
\cite{bau84}
since it includes the interaction $V_{bc}$ to all orders.

Recently, an alternative  theory of the Coulomb breakup has been
developed within the adiabatic (AD) model \cite{tos98,jal97}. This
theory assumes (i) that one of the fragments (the valence
nucleon) is neutral so that the projectile interacts with the
target only through the Coulomb interaction $V_{bt}$ of the core
fragment and the target nucleus, and (ii) that the relative excitation
energy $E_{bc}$ of the b-c system is much smaller than the total incident
energy so that $E_{bc}$ can be replaced by the constant separation
energy of the fragments in the projectile ground state. Following (ii)
the continuum spectrum of the $b-c$ system is assumed to be degenerate 
with the ground state.  Under the above assumptions, the wave function
$\we$ is found \cite{jal97} in the form
\begin{eqnarray}
\Psi_a^{(+)AD}(\xi_a,{\bf r}_1,{\bf r}_i) & = &
 \Phi_a(\xi_a,\ro)e^{i\alpha\ak \cdot \ro}\cp_a(\ak,{\bf r})
\end{eqnarray}
It is clear that substitution of Eq.~(12) to Eq.~(3) will
lead to a
factored form (similar to Eq.~(11)) of the reduced breakup
amplitude.
However, one limitation of this procedure should be brought into
attention.
For larger values of $\ro$, the wave function $\Psi_a^{(+)AD}$ vanishes
due to the presence of the factor $\Phi_a(\ro)$, whereas there may still be
contributions to the breakup from this region.
It has been argued \cite{tos98} that, due the presence of the interaction
$V_{bc}(r_1)$, the post form breakup amplitude may not be
sensitive to the domain where $\Psi_a^{(+)AD}$ is
inaccurate. However, since
the wave functions for the relative motion of the fragments
for $\ell > 0$ values have a
large spatial extension,
the application of this model to such cases may test
the need for the non-adiabatic corrections to the
theory.

The reduced amplitude in the adiabatic model is given by,
\begin{eqnarray}
{\hat \ell}\beta^{AD}_{\ell m} & = & \left [
\int d\ro e^{-i(\cq - \alpha\ak).\ro}V_{bc} (\ro) u_\ell (\ro) Y_{\ell m}
({\hat r}_1) \right ] \nonumber \\
& \times & \left [ \int d\ri \chi^{(-)*}_b(\bq,{\bf \ri})
e^{-i\delta\cq \cdot {\bf \ri}}\cp_a(\ak,{\bf \ri})
\right ]
\end{eqnarray}
It is obvious that this amplitude differs from that of the
FRDWBA, Eq.~(11), only in
the form factor part (the first of the factors), which is evaluated here
at the momentum transfer of $(\cq - \alpha \ak)$.
Equation~(13)
can also be obtained in the DWBA model by making a local momentum
approximation to the Coulomb distorted wave in
the initial channel of a reaction and by evaluating the local
momentum at $R = \infty$ with the momentum direction being the
same as that of the projectile. In both of the theories,
the Coulomb interaction between the fragments $b$ and the target is
treated non-perturbatively. The adiabatic model does
not make the weak coupling approximation of the DWBA. However,
it necessarily requires one of the
fragments (in this case $c$) to be neutral. In contrast,
the FRDWBA model can, in principle, be applied to the cases
where
both of the fragments $b$ and $c$ are charged \cite{shyam85}.
Furthermore, calculation of the nuclear breakup in the adiabatic model
is not as comparatively trivial \cite{jal97,jmc97,ron98}, as it
is in the case of FRDWBA.

The triple differential cross section of the reaction is given by
\begin{eqnarray}
{{d^3\sigma}\over{dE_bd\Omega_bd\Omega_c}} & = &
{2\pi\over{\hbar v_a}}\rho(E_b,\Omega_b,\Omega_c)
\sum_{\ell m}|\beta_{\ell m}|^2,
\end{eqnarray}
where $\rho(E_b,\Omega_b,\Omega_c)$ is the appropriate
\cite{cha00,fuchs} three-body phase space factor.

On substituting the Coulomb
distorted waves,
\begin{eqnarray}
\cm_b(\bq,{\ri}) & = &
 e^{-\pi\eta_b/2}\Gamma(1 + i\eta_b) e^{-i\bq.\ri}
 {_1F_1(-i\eta_b, 1, i(k_b r_i + \bq.\ri))} \,  , \\
                                          \nonumber \\
\cp_a(\ak,\ri) & = &
 e^{-\pi\eta_a/2}\Gamma(1 + i\eta_a) e^{i\ak.\ri}
 {_1F_1(-i\eta_a, 1, i(k_a r_i - \ak.\ri))}      \, ,
\end{eqnarray}
into Eqs.\ (11) and (13), one gets for the triple differential
cross section:
\begin{eqnarray}
{{d^3\sigma}\over{dE_bd\Omega_bd\Omega_c}} = {2\pi\over{{\hbar}v_a}}
\rho(E_b,\Omega_b,\Omega_c)
{4\pi^2\eta_a\eta_b\over (e^{2\pi\eta_b}-1)(e^{2\pi\eta_a}-1)}|I|^2
4\pi\sum_{\ell} |Z_{\ell}|^2.
\end{eqnarray}
In Eqs.~(15--17), $\eta$'s are the Coulomb parameters for the
respective particles. In Eq.\ (17),
$I$ is the Bremsstrahlung integral \cite{nord} which can be
evaluated in the closed form:
\begin{eqnarray}
I &=& -i{\Big[}B(0){\Big(}{{dD}\over{dx}}{\Big)}_{x=0}(-\eta_a\eta_b)\fa  \nonumber \\
& + & {\Big(}{{dB}\over{dx}}{\Big)}_{x=0} {\fB} {\Big]} \, ,
\end{eqnarray}
where
\begin{eqnarray}
B(x) = {4\pi\over{k^{2(i\eta_a+i\eta_b+1)}}}
{\Big[}(k^2 - 2{\bf k}.\ak -2xk_a)^{i\eta_a}
(k^2 - 2{\bf k}.\bq -2xk_b)^{i\eta_b}{\Big]},
\end{eqnarray}
\begin{eqnarray}
D(x) = {2k^2(k_ak_b+\ak.\bq)-4({\bf k}.\ak+xk_a)({\bf k}.\bq+xk_b)\over
{(k^2 - 2{\bf k}.\ak -2xk_a)(k^2 - 2{\bf k}.\bq -2xk_b)}} \, ,
\end{eqnarray}
with
\begin{eqnarray}
{\bf k} &=& \ak - \bq -\delta\cq.
\end{eqnarray}
The factor $Z_{\ell}$ contains the projectile structure information
and is given by
\begin{eqnarray}
Z_{\ell} = \int dr_1 r^2_1 j_{\ell} (k_1 r_1)\vv u_{\ell} (r_1),
\end{eqnarray}
with $k_1 = |\gamma\cq - \alpha {\bf K}|$, and
$k_1 = |\cq - \alpha \ak|$ for the cases of FRDWBA and adiabatic model,
respectively.

The total pure Coulomb one-nucleon
removal cross section for a given
$n\ell j$ configuration of the valence nucleon is obtained
by integrating Eq.~(13) over angles and energy of
fragment $b$ and over angles of the valence nucleon. Here,
$n$ is the principal quantum number and $\ell$ and $j$
are as defined in Eq.~(5).

For calculating the total cross section into a given
core-fragment final-state, the projectile ground state is
described as having a
configuration in which a valence nucleon, with single particle quantum numbers
$n\ell j$ and an associated spectroscopic factor $C^2S$, is
coupled to a
specific core state designated with $j_b$ in Eq.~(5). The total
cross section
$\sigma_C$ is the sum \cite{nav98,tos99} of the cross sections
calculated with configurations
(having  non-vanishing spectroscopic factors)
corresponding to all the allowed values of the
channel spin $j$

\section{Results and Discussions}
\subsection{Excitation of the bound states of $^{10}$Be in the Coulomb
breakup of $^{11}$Be.}

The one-neutron removal reaction of the type
$^9$Be($^{11}$Be,$^{10}$Be$\gamma$)X has been recently
studied~\cite{aum00}
at the beam energy of 60 MeV/nucleon.
Partial cross sections have been measured for
four states of the core fragment
$^{10}$Be: 0$^+$, 2$^+$, 1$^-$, and 2$^-$.
The data were analyzed in terms of an eikonal model of the nuclear
breakup reactions \cite{bar96,tos99}, with the spectroscopic factors
taken from \cite{war92}. It has been concluded in this study that
about $22\%$ of the total partial cross section went into the
excited states, and that the ground state of $^{11}$Be
consists of an admixture of the 1$s$  and 0$d$ single particle neutron
configurations with the spectroscopic factors of 0.74 and 0.18,
respectively.

We have calculated the pure Coulomb partial cross sections $\sigma_C$ to
the four $^{10}$Be final states in the
$^{208}$Pb($^{11}$Be,$^{10}$Be$\gamma$)X reaction at the beam energy
of 60 MeV/nucleon. The ground (0$^+$) and excited (3.368 MeV) (2$^+$)
states were assumed to correspond to the configurations
[1$s_{1/2}\nu$ $\otimes$ 0$^+$($^{10}$Be)] and
[0$d_{5/2}\nu$ $\otimes$ 2$^+$($^{10}$Be)], respectively, where
$\nu$ represents a relative neutron state.
The corresponding $C^2S$ values for these two configurations were
taken \cite{war92} to be 0.74 and 0.20, respectively, i.e.\ the
same as those used in \cite{aum00}. The excited 1$^-$ (5.956 MeV)
and 2$^-$ (6.256 MeV) states were assumed to stem from the configurations
[0$p_{3/2}\nu$ $\otimes$ 1$^-$($^{10}$Be)] and
[0$p_{3/2}\nu$ $\otimes$ 2$^-$($^{10}$Be)], respectively,
with the corresponding $C^2S$ values of 0.69 and 0.58.
These states could, in principle,
also result from the stripping of a 1$p_{3/2}$ neutron from the
$^{10}$Be(0$^+$) core of the $^{11}$Be ground state, producing a
[1s$_{1/2}$ $\otimes$ $^9$Be$({{3}\over{2}})^-$]1$^-$ and
[1s$_{1/2}$ $\otimes$ $^9$Be$({{3}\over{2}})^-$]2$^-$ types of $^{10}$Be$^*$
core. In the nuclear breakup case, the cross sections to 1$^-$ and
2$^-$ states calculated with the latter configurations were
found \cite{tos99} to be about 10-15$\%$ smaller than those
obtained with the preceding ones. We have carried out our calculations
with the former configuration for these states.
The one-neutron separation energy for the ground state
of $^{11}$Be, with the configuration in which $^{10}$Be remains in
its ground state, is taken to be $S_n = 0.504$~MeV.  For an
excited state,
the respective separation energy (SE) is assumed to be the sum
of $S_n$ and the excitation energy of that state with respect to
the ground state.

In each case, the neutron single particle wave function is calculated in
a central Woods-Saxon well of radius 1.15 fm and diffuseness 0.50 fm. The
depth of this well is adjusted to reproduce the corresponding
value of SE. By this procedure the root mean square (rms) radius of
the ground state of $^{11}$Be comes out to be
2.91 fm for the assumed rms radius of the $^{10}$Be core
of 2.28 fm.

Our results for the partial cross sections are shown in Table I.
It is evident from this table that in the case of pure Coulomb breakup
of a projectile with a halo ground state,
most of the cross section goes to the ground state (0$^+$) of
the core. The sum of the partial
\begin{table}[here]
\begin{center}
\caption {Calculated partial cross sections to the final
states of $^{10}$Be in the Coulomb breakup of $^{11}$Be on the
$^{208}$Pb
target at the beam energy of 60 MeV/nucleon. $I^\pi$ represents the
spin and parity of the populated states of the $^{10}$Be core. }
\vspace{0.5cm}
\begin{tabular}{|c|c|c|c|c|c|}
I$^\pi$ & $E_x$ & $\ell$ & $C^2S$ & $\sigma_C^{FRDWBA}$ &
$C^2S \cdot \sigma_C^{FRDWBA}$ \\
         &(\footnotesize{MeV})  &        &  & (\footnotesize{mb} ) &
(\footnotesize{mb}) \\
\hline
0$^+$   & 0.0   & 0   & 0.74 & 1536.48 & 1137.00 \\
\hline
2$^+$   & 3.368 & 2   & 0.20 &    2.09 &    0.42  \\
1$^-$   & 5.956 & 1   & 0.69 &    2.45 &    1.69  \\
2$^-$   & 6.256 & 1   & 0.58 &    2.07 &    1.20  \\
        &       & sum &      &    6.69 &    3.31  \\
\end{tabular}
\end{center}
\end{table}
\noindent
\noindent
cross sections corresponding to
all the excited states is less than 1$\%$ of that
to the ground state.
This is in sharp contrast
to the observations made on lighter targets where partial cross sections
corresponding to all the excited states represent about 22$\%$ of the total.
While there are no experimental data on the core-excitation
reaction induced by $^{11}$Be on a heavy target in the vicinity
of $^{208}$Pb, the measurements \cite{gui00}
of the (a,b$\gamma$) type of reactions with
$^{14}$B projectile
on $^{197}$Au gold target at the beam energy of 60 MeV/nucleon may be used
to test our results. In this experiment, no core-excited transitions were
seen.  Therefore, this lends support to our finding
that in A(a,b$\gamma$)X type of reactions, involving projectiles
which have a predominant $s$-wave neutron halo ground state, 
transitions to the excited states of the core corresponding to
the non-zero $\ell$-values of the neutron-core relative motion, are quite
weak in the pure Coulomb breakup reaction as compared to those in the
nuclear breakup process.

The suppression of the cross sections to the higher states can be traced
back to the strong dependence of the Coulomb breakup cross sections on
SE.  The latter enters in the corresponding expressions
through the momentum ${\bf k}$ (see Eq. (21)).
As was shown in \cite{ban93}, the modulus square of the bremsstrahlung
integral $I$ rises very steeply as the $k$ approaches zero,
which happens
as SE goes to zero ($|I|^2$ is infinite for $k$ = 0).
At larger values of SE (i.e.\ larger $k$) the rate
of the drop of $|I|^2$ becomes less drastic. This is reminiscent of
the behavior of the virtual photon numbers
in the theory of Coulomb excitation (see e.g. \cite{bert88}).
The value of $k$ is very small for SE equal to
$S_n$ and larger for SE corresponding to excited states.
This explains the reduction in the partial cross-sections to the
excited 2$^+$ state of $^{10}$Be core as compared to that to
its ground state. This also explains why the cross sections to the
excited states do not differ much from each other.
It may be useful to recall that,
due to the centrifugal barrier, the breakup cross sections for
non-$s$-wave projectiles are lower than those for the $s$-wave ones.
In case of the nuclear breakup, the dependence of
the cross section on SE is comparatively weaker
\cite{angela,shy91,das99}. This could be understood from the fact that
nuclear breakup cross sections are sensitive to the $b-c$ relative
wave functions at shorter distances which do not change much with
changes in the value of SE.

It should be interesting to compare the calculated pure Coulomb
partial cross
section for the $^{197}$Au($^{14}$B,$^{13}$B(g.s))X reaction, with its
experimental value given in Ref.\ \cite{gui00}. We performed
our calculations
with the configurations [1$s_{1/2}\nu$ $\otimes$ ${{3}\over{2}}^-$($^{13}$B)]
and [0$d_{5/2}\nu$ $\otimes$ ${{3}\over{2}}^-$($^{13}$B)] for the
$^{14}$B ground state. The resulting
cross sections were summed up, after multiplying them with the
corresponding spectroscopic factors of 0.663 and 0.306 (taken from
\cite{war92}), respectively, to obtain a value of 401 mb for
the pure Coulomb partial cross section for this reaction. The
corresponding experimental value is 638 $\pm$ 45 mb. The difference
between the calculated pure Coulomb and experimental partial cross sections
suggests that the nuclear, and Coulomb-nuclear interference terms could
contribute up to 40-50$\%$ in this reaction. This is an interesting
finding which underlines the need for extending the FRDWBA
theory
to include the nuclear breakup effects.  It should be stated here, that
the partial Coulomb cross sections obtained within the adiabatic model are
only a few percent larger than those of the FRDWBA theory and
show similar characteristics to those in Table I.

The longitudinal momentum distributions (LMD) for each of the
$^{10}$Be
core state are displayed in Fig.\ 2.
The solid and dashed lines
represent
the results of the FRDWBA and adiabatic model, respectively. We
note
that, while for the ground state of the $^{10}$Be core, the results of
the two theories are almost identical, they differ quite a bit from each
other for the excited states. It is for the first time that such
big differences are seen between the predictions of the two theories
for the momentum distributions.
\begin{figure}[here]
\begin{center}
\mbox{\epsfig{file=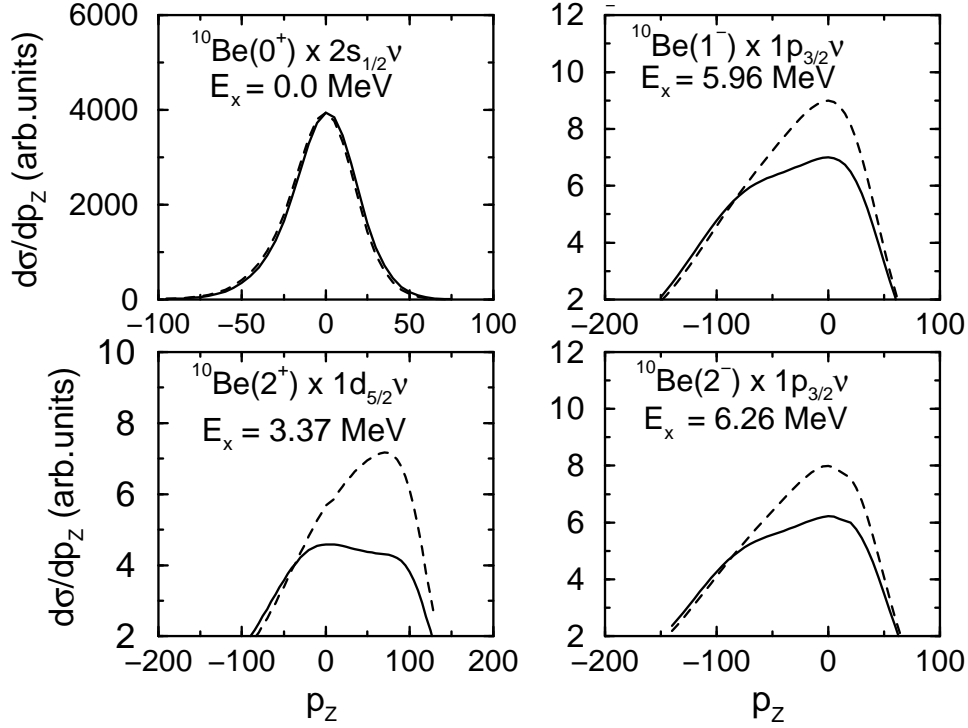,height=10.0cm}}
\end{center}
\caption {
Partial longitudinal momentum distributions
for the indicated states of $^{10}$Be fragment in the pure
Coulomb
one neutron removal reaction of $^{11}$Be on a $^{208}$Pb target at
the beam energy of 60 MeV/nucleon. The solid and dashed lines represent
the results obtained within FRDWBA and adiabatic models respectively.
The core - valence neutron configuration considered for each state is
indicated in the respective boxes.
}
\label{fig:figb}
\end{figure}
\noindent
Although due to unavailability of the experimental data for these cases,
it would be premature to comment upon suitability of either theory for
these excited states, a few speculative remarks can still be made. It is
not unreasonable to think that the adiabatic assumption (as discussed in the
previous section) may come under severe pressure for the excited states.
\begin{table}[here]
\begin{center}
\caption {Calculated partial cross sections to the final
states of $^{18}$C in the Coulomb breakup of $^{19}$C on a $^{208}$Pb
target at the beam energy of 60 MeV/nucleon.}
\vspace{0.5cm}
\begin{tabular}{|c|c|c|c|c|c|}
I$^\pi$ &$E_x$ & $\ell$ & $C^2S$ & $\sigma_C$ &
$C^2S$$\cdot$$\sigma_C$ \\
         &(\footnotesize{MeV})  &        &  & (\footnotesize{mb} ) &
(\footnotesize{mb})\\
\hline
0$^+$       & 0.0   & 0 & 0.58 & 993.2  & 576.1  \\
\hline
2$^+$       & 1.6   & 2   & 0.48 &   8.80 &   4.22 \\
0$^+$       & 4.0   & 0   & 0.32 &  13.38 &   4.28 \\
2$^+$,3$^+$ & 4.9   & 2   & 2.44 &   1.08 &   2.87 \\
            &       & sum &      &  23.26 &  11.37 \\
\end{tabular}
\end{center}
\end{table}
\noindent
Due to their non-$s$-wave nature,
the wave functions for the excited-state neutron-core motion
peak at larger values in the $r$-space yielding possibly a
significance to
the regime where the asymptotic form of the adiabatic wave
function [Eq. (12)] becomes inadequate.  It
would, therefore, be interesting to investigate the
importance of the non-adiabatic corrections \cite{ron98} to the
theory, for these cases.

The full width at half maximum (FWHM) of the calculated LMD for
the ground state of $^{10}$Be is 44 MeV/c which is consistent with
the experimental value of (47.5 $\pm$ 6) MeV/c seen in the
measurements on a $^9$Be target \cite{aum00}. This reconfirms that
LMDs are independent of the reaction mechanism and provide a very clean
way of determining the existence of halo structure in nuclei. The LMDs for
the excited states of $^{10}$Be are broad which is also consistent with
the observations made in \cite{aum00}. This indicates that the
respective states have a non-halo structure.

\subsection{Excitation of the bound states of $^{18}$C in the Coulomb
breakup of $^{19}$C.}

Table II displays results of
our calculations of the pure Coulomb partial cross sections
$\sigma_C$
for transitions to ground and three excited  bound states 
of $^{18}$C core 
in the $^{208}$Pb($^{19}$C,$^{18}$C$\gamma$)X reaction
at the beam energy of 60 MeV/nucleon. These states
have recently been seen \cite{val00}
\begin{figure}
\begin{center}
\mbox{\epsfig{file=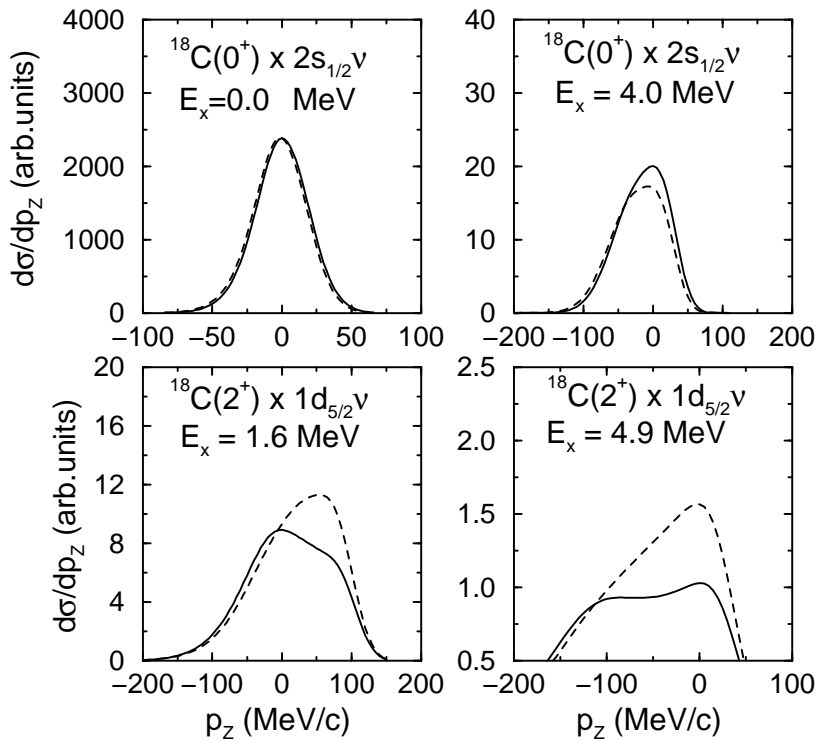,height=10.0cm}}
\end{center}
\caption {
Partial longitudinal momentum distributions
for the indicated states of $^{18}$C fragment in the pure
Coulomb one neutron removal
reaction of $^{19}$C on a $^{208}$Pb target, at the beam energy
of 60 MeV/nucleon. The solid and dashed lines represent the results
obtained within the FRDWBA and adiabatic models, respectively.
The core - valence
neutron configuration considered for each state is indicated in
the respective boxes.
}
\label{fig:figc}
\end{figure}
\noindent
in the $^9$Be($^{19}$C,$^{18}$C$\gamma$)X reaction at the
same beam energy.  The states of the $^{18}$C core
(with excitation energies of 0.0 MeV, 1.6 MeV, 4.0 MeV and 4.9 MeV)
are assumed to have the configurations,
[1$s_{1/2}\nu$ $\otimes$ 0$^+$($^{18}$C)],
[0$d_{5/2}\nu$ $\otimes$ 2$^+$($^{18}$C)],
[1$s_{1/2}\nu$ $\otimes$ 0$^+$($^{18}$C)], and
[0$d_{5/2}\nu$ $\otimes$ I$^\pi$($^{18}$C)], respectively.
The corresponding $C^2S$ values were taken
\cite{war92} to be 0.58, 0.48, 0.32
and 2.44, respectively, which are
the same as those used in \cite{val00}.
The value of $S_n$ for the ground state was taken to be 0.530 MeV.
We see that in this case too the ground state of $^{18}$C is
predominantly excited. The partial cross sections to
the excited states are somewhat larger than those seen in the case of
$^{10}$Be, since the excited $0^+$ state of $^{18}$C core can
have an $s$-wave
neutron relative motion. Yet, these contributions represent no more than
about 2$\%$ of the cross section to the ground state.

The LMD for each of the $^{18}$C
core states is shown in Fig.~3. The solid and dashed lines show
the results of the FRDWBA and adiabatic models, respectively.
In this case too we note that the predictions of the two models
differ
for the excited states of the core, while for the
ground state they agree very well with each other.
\begin{figure}[here]
\begin{center}
\mbox{\epsfig{file=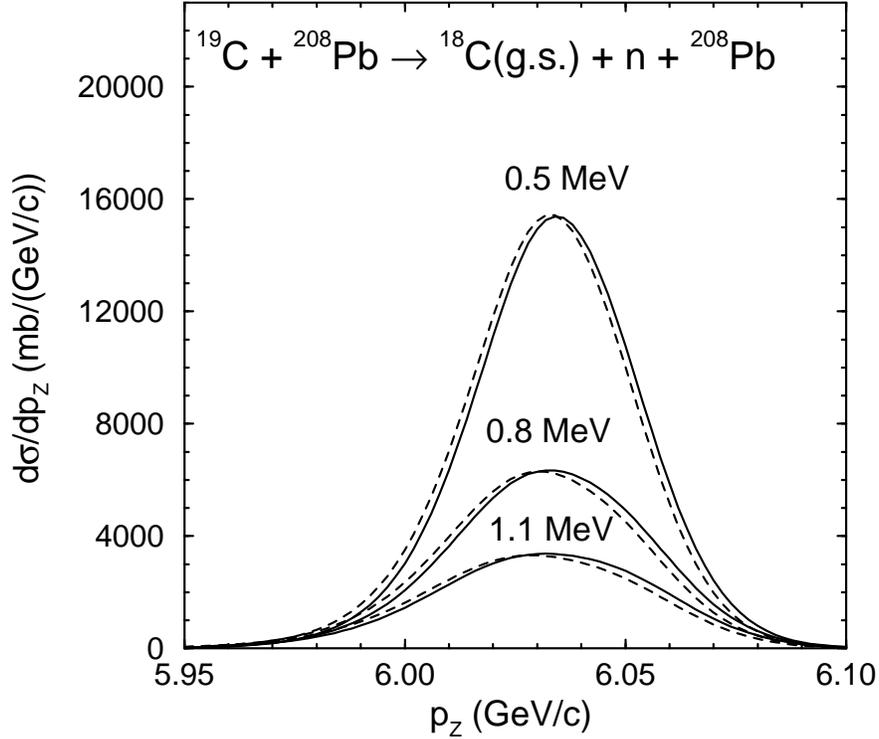,height=10.0cm}}
\end{center}
\vskip .3in
\caption{
Partial longitudinal momentum distribution for
the ground state of $^{18}$C in the pure Coulomb one neutron
removal reaction of $^{19}$C on a $^{208}$Pb target at the beam
energy of 60 MeV/nucleon for the core - valence neutron separation
energies of 0.5 MeV, 0.8 MeV and 1.1 MeV, as indicated. The solid
and dashed curves represent the results of the FRDWBA and adiabatic
models, respectively, in each case.
}
\label{fig:figd}
\end{figure}
\noindent
The value of $S_n$ for $^{19}$C is still an unsettled issue. The weighted
average of the atomic mass measurements carried out at Los Alamos and GANIL
\cite{wou88,orr91} suggests a value of 0.16 $\pm$ 0.11 MeV.
However, from the analysis \cite{nak99,ban00} of the data  on
the Coulomb dissociation of $^{19}$C, a higher value of 0.530 MeV
has been extracted. The interpretation of the recent data \cite{val00} on the
$^9$Be($^{19}$C,$^{18}$C$\gamma$)X reaction also suggests a higher value
of 0.8$\pm$0.3 MeV. Obviously, any conclusion drawn from the breakup
data strongly depends on the reaction mechanism and on the theory used for
the calculation of the breakup cross sections. With this precaution, we would
like to show here that the pure Coulomb breakup has some advantages over the
nuclear breakup process in this regard.

It should be mentioned here that one of the reasons for the uncertainty
in the value of $S_n$ in Ref.~\cite{val00} is the fact that,
due to the low
beam intensity in this experiment, the statistical errors associated with
the measured LMD for the ground state of $^{18}$C are large. The data
do not allow to distinguish between the nuclear breakup calculations of
the LMD done within the range
$S_n$ = 800 $\pm$ 300 keV. The difference in the peak
value of the nuclear LMD \cite{val00} calculated with $S_n$ = 1100 keV
and 500 keV is only about 1.6. In contrast,
the peak values of
the corresponding pure Coulomb LMD calculated with the same values of $S_n$
differ by a factor of about 4, as can be seen in Fig.~4.
This result is unlikely to be altered by the presence of the nuclear
breakup effects, as they tend to show up in the tail regions of the
LMDs. Thus A($^{19}$C,$^{18}$C(g.s))X type of reactions on a
heavy target may offer a better chance to put more definite
constraint on the value of $S_n$ for $^{19}$C.
\begin{figure}
\begin{center}
\mbox{\epsfig{file=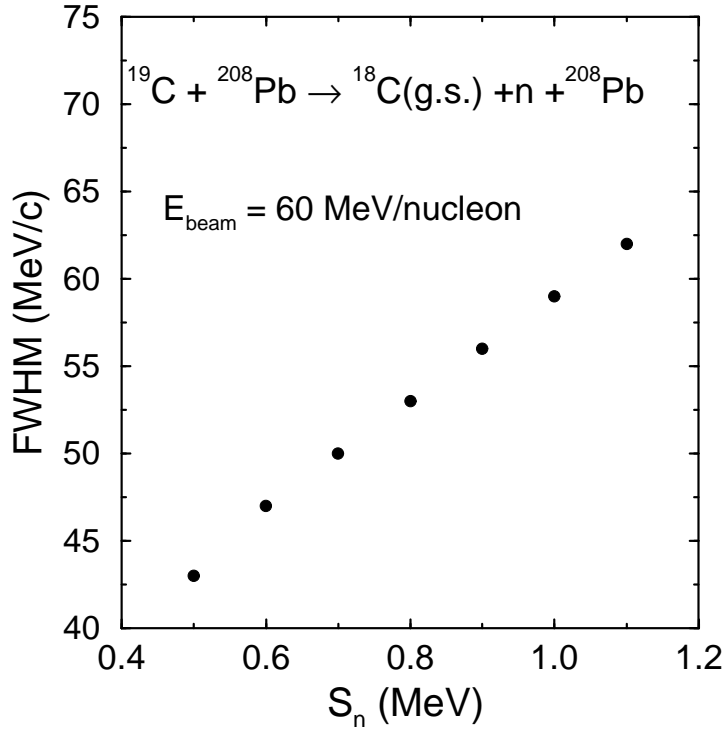,height=10.0cm}}
\end{center}
\vskip .3in
\caption {
Full width at half maximum (FWHM) of the longitudinal momentum
distribution of $^{18}$C(g.s) (shown by solid circles)
as a function of the core - valence neutron
separation energy in the same reaction
as in Fig. 4.
The FRDWBA and adiabatic model results are indistinguishable from each other.
}
\label{fig:fige}
\end{figure}
\noindent
Further insight into the value of $S_n$ from these reactions
can be
obtained from the full width at half maximum (FWHM) of the LMD.
In Fig.~5, we show the $S_n$
dependence of the FWHM of the LMD for the $^{18}$C(g.s) in the
same reaction as in Fig.~4. It can be seen that FWHM increases
from
42 MeV/c to about 65 MeV/c as $S_n$ increases from 500 keV to 1100 keV.
The variation of the corresponding FWHM in the nuclear breakup
case is relatively weaker.

\section{Summary and Conclusions}

In this paper we calculated the pure Coulomb breakup
contributions to the partial cross sections and to the longitudinal
momentum distributions for the ground and excited states of the core
fragments observed in $^{208}$Pb($^{11}$Be,$^{10}$Be$\gamma$)X and
$^{208}$Pb($^{19}$C,$^{18}$C$\gamma$)X types of one-neutron
removal reactions, at the beam energy of 60 MeV/A. These reactions have
recently been studied at the Michigan State University but on
the light $^9$Be target; hence, these data are dominated by the nuclear
breakup effects. One of our aims was to see in what way the Coulomb
dominated reaction mechanism was different and could
supplement the conclusions derived from the pure nuclear breakup
studies of the nuclei. The advantage of the pure Coulomb break
up process is that the corresponding theory has no free
adjustable parameter, and assuming that the mutual excitation of the
target by the Coulomb force is negligible,
the inelastic breakup mode does not contribute to this process.

As in the previous studies \cite{nav00,aum00,val00},
we assumed that the coupling between the core states
is weak and that there is no dynamical excitation of these states.
Thus, the reaction can only populate those core states which
have a non-zero spectroscopic factor for a given neutron-core configuration
in the projectile ground state. We employed both the finite
range DWBA and adiabatic model of the Coulomb breakup theory in our
calculations. In earlier studies \cite{cha00} of the inclusive Coulomb
dissociation cross sections, the two theories produced nearly identical
results for the momentum distributions of heavy fragments.

We found that in reactions of the type A(a,b$\gamma$)X on a
heavy target, the core ground state is predominantly excited; higher
energy states account for only a few percent of the total cross section.
This finding is in contrast to the results
obtained on similar reactions on a light target, where about
a quarter of the total breakup cross section could be due to transitions
to core excited states.  Our finding
is supported by a recent measurement \cite{gui00} of the
$^{197}$Au($^{14}B$,$^{13}$B$\gamma$)X  reaction at the beam energy of
60 MeV. The reason for this
difference is attributed to the fact that pure Coulomb breakup
cross sections drop very strongly as the separation energy increases.
On the other hand, the nuclear breakup cross sections decrease slowly
with increasing SE. Therefore, such reactions on a heavy target are
potentially a more useful tool for investigating the properties of the ground
state of the core fragments.

A rather interesting result of our study is that the
finite-range DWBA and the adiabatic theories of Coulomb
breakup
lead to very different longitudinal momentum distributions for the
excited states of the core fragments. It is probably for the
first time that such a large difference is seen in the
predictions of
two theories for the momentum
distributions. This should provide
some impetus to look for the non-adiabatic corrections to the
adiabatic approximation which may come under some pressure for
the excited states.

Coulomb dominated breakup reactions may provide a better way for
resolving the uncertainty associated with the one neutron
separation energy of $^{19}$C. The peak value and the full
width
at half maximum of the longitudinal momentum distributions for
the ground
state of $^{18}$C core are more sensitive to the one-neutron
separation energy in the Coulomb breakup process than
in the nuclear breakup. In the latter case, the dependence could be so
weak that the data with limited statistics may not
allow to
distinguish between the values of these quantities calculated with quite
different one-neutron separation energies.

One of the authors (RS) would like to acknowledge several useful discussions
with Gregers Hansen. This work was supported by the National Science
Foundation under Grant PHY-0070818.

\end{document}